\documentclass[aps,prl,twocolumn,groupedaddress,showpacs,showkeys]{revtex4-1}
\usepackage{amsfonts}
\usepackage{mathtools}
\usepackage{amsmath}
\usepackage{amssymb}
\usepackage{float}
\usepackage{subdepth}
\usepackage{color}
\usepackage{graphicx}
\usepackage{natbib}
\usepackage{hyperref}
\usepackage{url}
\usepackage{soul}
\hypersetup{colorlinks, linkcolor={blue}, citecolor={blue}, urlcolor={blue}}
\begin{document}
\title{Excitations of the ferroelectric order}
\author{Ping Tang$^{1}$}
\author{Ryo Iguchi$^{2}$}
\author{Ken-ichi Uchida$^{2,3,4}$}
\author{Gerrit E. W. Bauer$^{1,3,4,5}$}
\affiliation{$^1$WPI-AIMR, Tohoku
University, 2-1-1 Katahira, 980-8577 Sendai, Japan}
\affiliation{$^2$National Institute for Materials Science, Tsukuba 305-0047, Japan}
\affiliation{$^3$Institute for Materials Research, Tohoku University, 2-1-1 Katahira, 980-8577 Sendai, Japan}
\affiliation{$^4$Center for Spintronics Research Network, Tohoku University, Sendai 980-8577, Japan}
\affiliation{$^5$Zernike Institute for Advanced Materials, University of Groningen, 9747 AG Groningen, Netherlands}
\begin{abstract}
We identify the bosonic excitations in ferroelectrics that carry electric
dipoles from the phenomenological Landau-Ginzburg-Devonshire theory. The
\textquotedblleft ferron\textquotedblright\ quasi-particles emerge from the
concerted action of anharmonicity and broken inversion symmetry. In contrast
to magnons, the transverse excitations of the magnetic order, the ferrons in
displacive ferroelectrics are longitudinal with respect to the ferroelectric
order. Based on the ferron spectrum, we predict temperature dependent
pyroelectric and electrocaloric properties, electric-field-tunable heat and
polarization transport, and ferron-photon hybridization.
\end{abstract}
\maketitle
The spontaneous emergence of order in condensed matter below a critical
temperature breaks a symmetry, while the low-energy collective excitations of
the order parameter tend to restore it. The latter can often be modeled by
non-interacting quasi-particles that in extended system are plane waves with a
well-defined dispersion relation. Their lifetime is finite due to
self-interactions or coupling with the environment. Wave packets of these
quasiparticles transport energy, momentum, and order parameter with the group
velocity from the dispersion relations.

Lattice vibrations disturb the translational symmetry of homogeneous elastic
media, and phonons are the associated quasi-particles. The excitations of a
magnetic order are spin waves. The associated quanta, the magnons, carry
magnetic moments that reduce the magnetization and can transport spin angular
momentum and energy \cite{kruglyak2010magnonics,chumak2015magnon}. Gradients
of temperature and magnon chemical potential
\cite{cornelissen2015long,cornelissen2016magnon} induce magnon spin and heat
currents, with associated spin Seebeck \cite{uchida2010spin} and spin Peltier
\cite{flipse2014observation,daimon2016thermal} effects.

Ferroelectric materials exhibit ordered electric dipoles with unique
dielectric, pyroelectric, piezoelectric and electrocaloric properties
\cite{xu2013ferroelectric}, with many analogies with ferromagnets
\cite{spaldin2007analogies}. However, to the best of our knowledge, the
quasi-particles associated to the ferroelectric order have so far remained
elusive. We previously addressed the elementary excitations of ferroelectrics
or \textquotedblleft ferrons\textquotedblright\ and the associated
polarization and heat transport \cite{bauer2021theory,tang2021ferroelectric}
by a phenomenological diffusion equation and a simple ball-spring model. The
latter was inspired by magnons, which are transverse fluctuations that
preserve the magnitude of the local magnetization. The assumption of local
electric dipoles with fixed modulus should hold for order-disorder
ferroelectrics such as NaNO$_{2}$ that are formed by stable molecular dipoles
\cite{blinc1972dynamics}. However, most ferroelectrics are \textquotedblleft
displacive\textquotedblright, i.e., formed by the condensation of a particular
soft phonon \cite{cochran1959crystal, cochran1960crystal} with a flexible
dipole moment (or are of mixed type
\cite{muller1982displacive,dalal1998coexistence,zalar2003nmr,
bussmann2009precursor}), and cannot be described by our previous model.

\begin{figure}[ptb]
\centering
\par
\includegraphics[width=7.2cm]{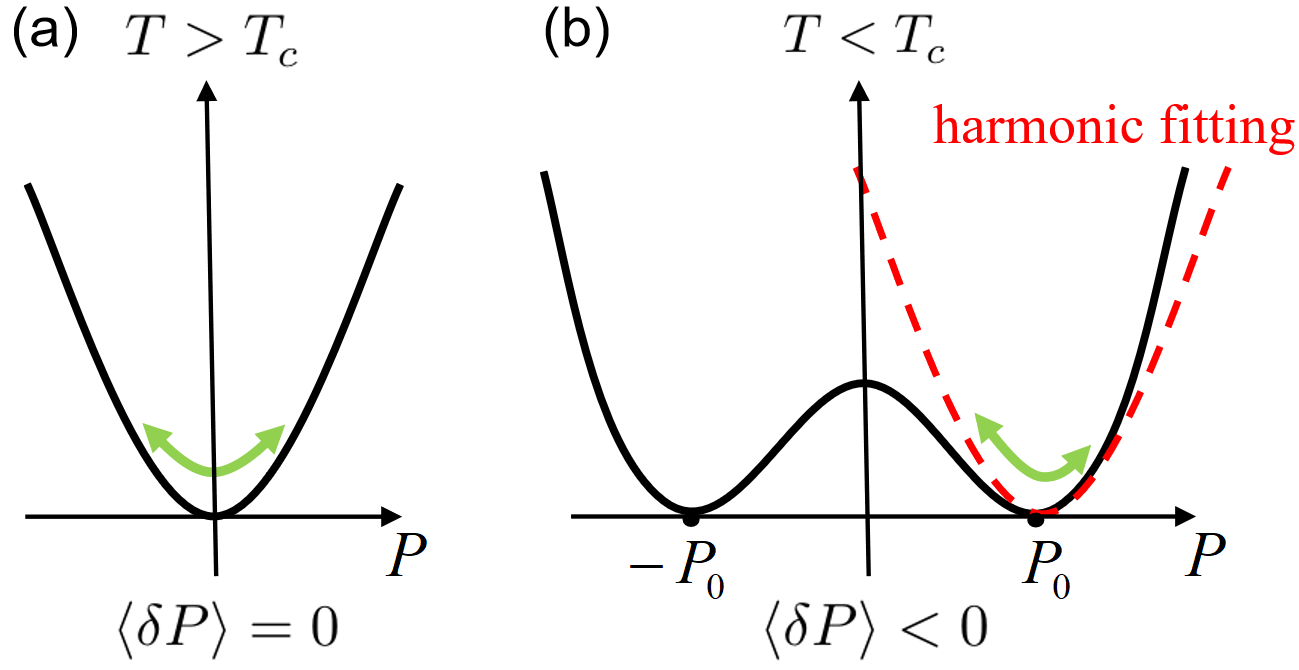}\newline\caption{Landau potential
energy landscape for polarization fluctuations in ferroelectrics (green
arrows). (a) Above the critical temperature $T_{c}$ in the paraelectric phase
the potential is symmetric for the fluctuations around the minimum $P_{0}=0$
and the average of the fluctuations $\langle\delta P\rangle=\langle
P-P_{0}\rangle=0$ even in the presence of anharmonicity (see Eq.~(\ref{anflu}%
)). (b) Below $T_{c}$ the ferroelectric order breaks inversion symmetry and
polarization fluctuates around finite $\pm P_{0}$ (e.g., the positive one in
the figure) in an asymmetric potential, therefore carrying a non-vanishing
average electric dipoles, i.e., $\langle\delta P\rangle=\langle P-P_{0}%
\rangle<0$. }%
\label{Fig-1}%
\end{figure}

In this Letter, we formulate the quasi-particle excitations of displacive
ferroelectrics in the framework of the Landau-Ginzburg-Devonshire (LGD) theory
\cite{devonshire1949xcvi,devonshire1951cix}, which has been widely used to
model ferroelectrics over a broad temperature range \cite{salje1990phase}.
These ferrons are longitudinal rather than transverse fluctuations and carry
electric polarization because of the non-parabolicity of the free energy
around the local minima below the phase transition (see Fig.~\ref{Fig-1}). The
parameters of the LGD free energy are well-known for many materials, which
allows quantitative predictions of their thermodynamic and transport properties.

\emph{Model:} The LGD free energy $F(\mathbf{P})$ for a ferroelectric is a
functional of the macroscopic polarization texture $\mathbf{P}(\mathbf{r})$
that obeys the crystal symmetry of the parent paraelectric phase
\cite{cao2008constructing}. For a uniaxial ferroelectric formed out of a
centrosymmetric paraelectric phase the (Gibbs) free energy is an integral over
the sample volume $V$ \cite{chandra2007landau}:
\begin{equation}
F=\int d^{3}\mathbf{r}\left(  \frac{g}{2}(\nabla\mathbf{P})^{2}+\frac{\alpha
}{2}P^{2}+\frac{\beta}{4}P^{4}+\frac{\lambda}{6}P^{6}-\mathbf{E}%
\cdot\mathbf{P}\right)  ,\label{free}%
\end{equation}
where $\alpha$, $\beta$ and $\lambda>0$ are the Landau coefficients, $g>0$ is
the Ginzburg-type parameter that accounts for the energy cost of polarization
textures and $\mathbf{E}$ an external electric field. A constant spontaneous
polarization $(\mathbf{P}_{0})$ minimizes $F(\mathbf{P}) $ of a
uniform\ medium when
\begin{equation}
\alpha P_{0}+\beta P_{0}^{3}+\lambda P_{0}^{5}=E\label{spon}%
\end{equation}
where $P_{0}$ ($E$) is the modulus of the vectors $\mathbf{P}_{0}$
($\mathbf{E}$) and $\mathbf{E\Vert P}_{0}.$ Below a critical temperature
$T_{c}$ the system orders in a first (second)-order phase transition for
$\beta<0$ ($\beta>0$) with $P_{0}^{2}=\left(  -\beta+\sqrt{\beta^{2}%
-4\alpha\lambda}\right)  /2\lambda$ for $E=0$.

In the presence of fluctuations, the longitudinal polarization dynamics
($\mathbf{P}\Vert\mathbf{P}_{0}$) obeys the Landau-Khalatnikov-Tani equation
\cite{tani1969dynamics,ishibashi1989phenomenological,sivasubramanian2004physical,widom2010resonance}%
,
\begin{equation}
m_{p}\frac{\partial^{2}P}{\partial t^{2}}+\gamma\frac{\partial P}{\partial
t}=-\frac{\partial F}{\partial P}+E_{\mathrm{th}},\label{LKT}%
\end{equation}
where $m_{p}=(\varepsilon_{0}\omega_{p}^{2})^{-1}$ is the polarization inertia,
$\varepsilon_{0}$ the vacuum dielectric constant, and $\gamma$ a
phenomenological damping constant. The plasma frequency $\omega_{p}$ depends
on the ionic masses $M_{i}$ and charges $Q_{i}$ in the unit cell of volume
$V_{0}$ as $\omega_{p}^{2}=(\varepsilon_{0}V_{0})^{-1}\sum_{i}Q_{i}^{2}/M_{i}$
\cite{sivasubramanian2004physical}. $E_{\mathrm{th}}(\mathbf{r},t)$ is a
Langevin noise field that obeys a fluctuation-dissipation theorem
\cite{Landau1980},
\begin{align}
\langle E_{\mathrm{th}}(\mathbf{q},\omega) &  E_{\mathrm{th}}^{\ast
}(\mathbf{q}^{\prime},\omega^{\prime})\rangle=\frac{(2\pi)^{4}\gamma
\hbar\omega\delta(\mathbf{q}-\mathbf{q}^{\prime})\delta(\omega-\omega^{\prime
})}{\tanh(\hbar\omega/2k_{B}T)}\nonumber\label{Langevin}\\
&  \overset{k_{B}T\gg\hbar\omega}{\rightarrow}(2\pi)^{4}2\gamma k_{B}%
T\delta(\mathbf{q}-\mathbf{q}^{\prime})\delta(\omega-\omega^{\prime}),
\end{align}
where $\langle\cdots\rangle$ is an ensemble average, $E_{\mathrm{th}%
}(\mathbf{q},\omega)=\int dt\int d^{3}\mathbf{r}E_{\mathrm{th}}(\mathbf{r}%
,t)e^{-i\mathbf{q}\cdot\mathbf{r}+i\omega t}$ the Fourier component of
$E_{\mathrm{th}}(\mathbf{r},t)$ and the second line indicates the classical
white noise limit. Substituting the small fluctuations $\delta P(\mathbf{r}%
,t)=P(\mathbf{r},t)-P_{0}$ into Eq.~(\ref{LKT}):%
\begin{equation}
\hat{G}^{-1}\delta P=E_{\mathrm{th}}-\left(  3\beta+10\lambda P_{0}%
^{2}\right)  P_{0}\delta P^{2}+\mathcal{O}(\delta P^{3})\label{flu1}%
\end{equation}
where $\hat{G}^{-1}\equiv m_{p}\partial_{t}^{2}+\gamma\partial_{t}-g\nabla
^{2}+(\alpha+3\beta P_{0}^{2}+5\lambda P_{0}^{4})$ is an inverse propagator.
The non-linear terms on the right-hand side of Eq.~(\ref{flu1}) are
proportional to the anharmonicity parameters $\beta$ and $\lambda$ in
Eq.~(\ref{free}). At temperatures sufficiently below the $T_{c}$ the
fluctuations $E_{\mathrm{th}}$ are small and we may solve Eq.~(\ref{flu1})
iteratively. To leading order,
\begin{equation}
\delta P=\delta P_{\mathrm{h}}-\left(  3\beta+10\lambda P_{0}^{2}\right)
P_{0}\hat{G}\delta P_{\mathrm{h}}^{2}+\mathcal{O}(E_{\mathrm{th}}%
^{3})\label{flu2}%
\end{equation}
where $\delta P_{\mathrm{h}}\equiv\hat{G}E_{\mathrm{th}}$ are the harmonic
thermal fluctuations that on average do not change the polarization since
$\left\langle \delta P_{\mathrm{h}}\right\rangle =0$. In Fourier space
\begin{equation}
\delta P_{\mathrm{h}}(\mathbf{q},\omega)=\frac{E_{\mathrm{th}}(\mathbf{q}%
,\omega)}{m_{p}(\omega_{\mathbf{q}}^{2}-\omega^{2})-i\omega\gamma}\label{hfl}%
\end{equation}
where $\omega_{\mathbf{q}}=m_{p}^{-1/2}(\alpha+3\beta P_{0}^{2}+5\lambda
P_{0}^{4}+g\mathbf{q}^{2})^{1/2}$ is the dispersion relation. Assuming weak
dissipation $\gamma\ll m_{p}\omega_{\mathbf{q}}$, Eqs.~(\ref{Langevin}),
(\ref{flu2}) and (\ref{hfl}) leads to fluctuations
\begin{equation}
\langle\delta P\rangle=-\frac{\hbar(3\beta+10\lambda P_{0}^{2})P_{0}}%
{2m_{p}(\alpha+3\beta P_{0}^{2}+5\lambda P_{0}^{4})}\int\frac{d^{3}\mathbf{q}%
}{(2\pi)^{3}}\frac{1}{\omega_{\mathbf{q}}}\coth\frac{\hbar\omega_{\mathbf{q}}%
}{2k_{B}T}\label{anflu}%
\end{equation}
that suppress the ground state polarization $P_{0}$ because of the
anharmonicity, see Fig.~\ref{Fig-1}. We may quantize the harmonic fluctuations
as
\begin{equation}
\delta\hat{P}_{\mathrm{h}}=\sqrt{\frac{\hbar}{2m_{p} V}}\sum_{\mathbf{q}}%
\hat{a}_{\mathbf{q}}\frac{e^{i\mathbf{q}\cdot\mathbf{r}}}{\sqrt{\omega
_{\mathbf{q}}}}+\mathrm{H.c}{\normalsize .}\label{QF}%
\end{equation}
where $\hat{a}_{\mathbf{q}}$ ($\hat{a}_{\mathbf{q}}^{\dagger}$) represents the
bosonic annihilation (creation) operator of \textquotedblleft
ferrons\textquotedblright\ with wave vector $\mathbf{q}$ and frequency
$\omega_{\mathbf{q}}$. After substracting the zero-point fluctuations, the
elementary electric dipole carried by a single ferron is $\delta
p_{\mathbf{q}}=\langle\mathbf{q}|\delta\hat{P}|\mathbf{q}\rangle
-\langle0|\delta\hat{P}|0\rangle$, where $|\mathbf{q}\rangle=\hat
{a}_{\mathbf{q}}^{\dagger}|0\rangle$ and $|0\rangle$ the vacuum. By
substituting Eq.~(\ref{QF}) into Eq.~(\ref{flu2}),
\begin{equation}
\delta p_{\mathbf{q}}=-\frac{\hbar(3\beta+10\lambda P_{0}^{2})P_{0}}%
{m_{p}(\alpha+3\beta P_{0}^{2}+5\lambda P_{0}^{4})}\frac{1}{\omega_{\mathbf{q}%
}}.\label{emp}%
\end{equation}
Using the non-linear dielectric susceptibility $\chi=\partial P_{0}/\partial
E=(\alpha+3\beta P_{0}^{2}+5\lambda P_{0}^{4})^{-1}$ that follows from
Eq.~(\ref{spon}), Eq.~(\ref{emp}) can be rewritten as
\begin{equation}
\delta p_{\mathbf{q}}=\frac{\hbar}{2m_{p}}\frac{\partial\ln\chi}{\partial
P_{0}}\frac{1}{\omega_{\mathbf{q}}}.\label{emp1}%
\end{equation}
Eq. (\ref{emp}) and Eq.~(\ref{emp1}) agree with the intuitive relation
\begin{equation}
\delta p_{\mathbf{q}}=-\frac{\partial\hbar\omega_{\mathbf{q}}}{\partial
E}\label{emp2}%
\end{equation}
which also holds for $E\neq0$. According to Eq.~(\ref{emp}) the ferron
electric dipole reduces $\mathbf{P}_{0}$ (i.e., $\partial\ln\chi/\partial
P_{0}<0$) and emerges from the anharmonicity of the free energy below the
phase transition. As in order-disorder ferroelectrics
\cite{bauer2021theory,tang2021ferroelectric} and in contrast to the magnetic
dipole associated to magnons, the electric dipole of the longitudinal ferrons
depends strongly and non-universally on the wave vector. In the paraelectric
phase, the spontaneous polarization vanishes and hence $\delta p_{\mathbf{q}%
}=0$, but a strong enough applied external field polarizes the paraelectric
state and its elementary excitations as well.

The expansion to leading order in the amplitudes limits quantitative
predictions to temperatures sufficiently below $T_{c}$. However, we may profit
in the future from the large knowledge base on computing phononic
non-linearities in complex materials \cite{tadano2018first}.

We assume dominance of a single-band soft mode that triggers the
symmetry-breaking structural phase transitions to the ferroelectric state. In
displacive ferroelectrics this is the lowest soft optical phonon that vibrates
parallel to $\mathbf{P}_{0}$. Hybridization with other, such as acoustic,
phonon modes can become significant for some physical properties \cite{PT}.

The free energy Eq. (\ref{free}) does not introduce non-parabolicities to the
transverse oscillations, which therefore do not carry any dipolar moment.
Order-disorder ferroelectrics can also be treated by Landau theory, but
polarized fluctuations only emerge by introducing non-linearities in the
transverse amplitudes. At sufficiently low temperatures this can conveniently
be achieved by the constraint $\left\vert \mathbf{P}\right\vert =P_{0}$, which
to leading order gives rise to a finite dipole of the transverse ferrons,
analogous to the magnetic moment of magnons
\cite{bauer2021theory,tang2021ferroelectric}. Longitudinal and
transverse\ ferrons\ may coexist in some multiaxial materials.

Since the LGD parameters are well documented for many ferroelectric materials
\cite{haun1987thermodynamic,pertsev1998effect,
scrymgeour2005phenomenological,li2005phenomenological,
hlinka2006phenomenological,liang2009thermodynamic}, we are in an excellent
position to quantitatively study ferron-related thermodynamic, optical, and
transport properties. Table \ref{tab:table1} summarizes the key information
for selected displacive ferroelectrics with perovskite crystal structure at
room temperature.

\emph{Pyroelectricity and electrocalorics.} Pyroelectricity (electrocalorics)
is the change of polarization (entropy) under a temperature (electric field)
change \cite{whatmore1986pyroelectric,muralt2001micromachined,
mischenko2006giant,neese2008large,li2020pyroelectric}. They are conventionally
calculated directly by the LGD free energy with linear temperature dependence
of the Landau quadratic coefficient ($\alpha$)
\cite{muralt2001micromachined,li2020pyroelectric}. However, this approach is
valid only near the phase transition. At lower temperatures the fluctuations
are well represented by the ferrons, and $\alpha$ becomes temperature
independent. The total polarization is $P(T)=P(0)+\Delta P(T)$ with
\begin{align}
\Delta P(T) &  =\int\frac{d^{3}\mathbf{q}}{(2\pi)^{3}}f_{0}\left(
\xi_{\mathbf{q}}\right)  \delta p_{\mathbf{q}}\nonumber\\
&  \rightarrow-\frac{\hbar(3\beta+10\lambda P_{0}^{2})P_{0}}{(2\pi
g)^{3/2}m_{p}^{1/2}}\frac{1}{\xi_{0}^{3/2}}\exp\left(  -\xi_{0}\right)
,\label{lowp}%
\end{align}
where $f_{0}(\xi_{\mathbf{q}})=[\exp(\xi_{\mathbf{q}})-1]^{-1}$ is the Planck distribution, $\xi_{\mathbf{q}}=\hbar\omega_{\mathbf{q}}/k_{B}T$ and in the second step we
took the low temperature limit $\xi_{0}=\hbar\omega_{0}/k_{B}T\gg1$ with $\omega_{0}=m_{p}
^{-1/2}(\alpha+3\beta P_{0}^{2}+5\lambda P_{0}^{4})^{1/2}$ the ferron gap
(at $E=0$). By disregarding the temperature dependence of material parameters,
the low-temperature pyroelectric coefficient we arrive at the thermally
activated form
\begin{equation}
\frac{\partial\Delta P}{\partial T}\rightarrow-\frac{(\hbar k_{B}%
)^{1/2}(3\beta+10\lambda P_{0}^{2})P_{0}}{(2\pi g)^{3/2}(m_{p}\omega_{0}%
)^{1/2}}\frac{1}{\sqrt{T}}\exp\left(  -\xi_{0}\right)  .\label{pyr}%
\end{equation}
The electrocaloric coefficient, i.e. the isothermal entropy change with
electric field, is according to the Maxwell relation
\begin{equation}
\left(  \frac{\partial\Delta S}{\partial E}\right)  _{T}=\left(
\frac{\partial\Delta P}{\partial T}\right)  _{E}.
\end{equation}

The temperature dependence deviates strongly from a Curie-Weiss power-law.
Glass and Lines \cite{glass1976low} derived the scaling relation
Eq.~(\ref{pyr}) in order to explain the low-temperature pyroelectricity of
LiNbO$_{3}$ and LiTaO$_{3},$ thereby implicitly introducing the ferron concept
for equilibrium properties a long time ago. Lang et al.
\cite{lang1969pyroelectric} observed a negative pyroelectric coefficient in
BaTiO$_{3}$ ceramic at low temperature, whose absolute value increases
exponentially with temperature, in qualitative agreement with Eq.~(\ref{pyr}).
However, the experimental $\partial\Delta P/\partial T=-5\times10^{-7}%
$\thinspace C/(m$^{2}$K) at $4.9$\thinspace K is much larger than
Eq.~(\ref{pyr}), which has been ascribed to a polarization of acoustic phonons
coupled to the soft mode \cite{born1945quantum,szigeti1975temperature}%
.\textit{\ }

\textit{Polarization and heat transport by ferrons}. We consider here diffuse
and ballistic ferron transport in bulk ferroelectrics \cite{bauer2021theory}
and through constrictions \cite{tang2021ferroelectric}, respectively. In the
former case we focus on homogeneous single-domain ferroelectrics at local
thermal equilibrium with an electric field generated by internal polarization
and external charges. Electric field ($\partial E$) and temperature ($\partial
T$) gradients set along the $x$ direction induce polarization $\left(
j_{p}\right)  $ and heat $\left(  j_{q}\right)  $ current densities. The
driving forces include non-equilibrium contributions from polarization and
heat accumulations that should be computed self-consistently
\cite{bauer2021theory}. We can derive the polarization $\left(  \sigma\right)
$ and thermal $\left(  \kappa\right)  $ conductivities and the Seebeck $\left(
S_{d}\right)  $ and Peltier $\left(  \Pi_{d}\right)  $ coefficients in the
linear response relations
\begin{equation}
\left(
\begin{matrix}
-j_{p}\\
j_{q}%
\end{matrix}
\right)  =\sigma\left(
\begin{matrix}
1 & S_{d}\\
\Pi_{d} & \kappa/\sigma
\end{matrix}
\right)  \left(
\begin{matrix}
\partial E\\
-\partial T
\end{matrix}
\right)
\end{equation}
by the Landau theory introduced above. The linearized Boltzmann transport
equation of the ferron gas in a constant relaxation time approximation
\cite{bauer2022magnonics} yields
\begin{align}
\sigma &  =\frac{\tau}{\hbar}\int(v_{\mathbf{q}}^{x})^{2}(\delta
p_{\mathbf{q}})^{2}\left(  -\frac{\partial f_{0}}{\partial\omega_{\mathbf{q}}%
}\right)  \frac{d^{3}\mathbf{q}}{(2\pi)^{3}}\nonumber\\
&  =\frac{\tau\hbar}{8\pi^{2}m_{p}^{3/2}g^{1/2}}\left[  \frac{\partial\ln\chi
}{\partial P_{0}}\right]  ^{2}\left\{
\begin{array}
[c]{c}%
\sqrt{\frac{\pi}{2}}\xi_{0}^{-3/2}e^{-\xi_{0}},\\
\frac{\pi}{16}\xi_{0}^{-1},
\end{array}%
\begin{array}
[c]{c}%
\xi_{0}\gg1\\
\xi_{0}\ll1
\end{array}
\right. \\
S_{d} &  =\frac{\tau}{\hbar(\sigma T)}\int(v_{\mathbf{q}}^{x})^{2}(-\delta
p_{\mathbf{q}})\hbar\omega_{\mathbf{q}}\left(  -\frac{\partial f_{0}}%
{\partial\omega_{\mathbf{q}}}\right)  \frac{d^{3}\mathbf{q}}{(2\pi)^{3}%
}\nonumber\\
&  =\frac{\tau k_{B}^{2}T}{12\pi^{2}\hbar(m_{p} g)^{1/2}\sigma}\frac
{\partial\ln\chi}{\partial P_{0}}\left\{
\begin{array}
[c]{c}%
3\sqrt{\frac{\pi}{2}}\xi_{0}^{1/2}e^{-\xi_{0}},\\
\frac{\pi^{2}}{3},
\end{array}%
\begin{array}
[c]{c}%
\xi_{0}\gg1\\
\xi_{0}\ll1
\end{array}
\right. \\
\kappa &  =\frac{\tau}{\hbar T}\int(v_{\mathbf{q}}^{x})^{2}(\hbar
\omega_{\mathbf{q}})^{2}\left(  -\frac{\partial f_{0}}{\partial\omega
_{\mathbf{q}}}\right)  \frac{d^{3}\mathbf{q}}{(2\pi)^{3}}\nonumber\\
&  =\frac{\tau k_{B}^{4}T^{3}m_{p}^{1/2}}{6\pi^{2}\hbar^{3}g^{1/2}}\left\{
\begin{array}
[c]{c}%
3\sqrt{\frac{\pi}{2}}\xi_{0}^{5/2}e^{-\xi_{0}},\\
\frac{4\pi^{4}}{15},
\end{array}%
\begin{array}
[c]{c}%
\xi_{0}\gg1\\
\xi_{0}\ll1
\end{array}
\right. \label{DTR}%
\end{align}
and the Kelvin-Onsager relation $\Pi_{d}=TS_{d}$. Here $\tau$ is the ferron
relaxation time, $v_{\mathbf{q}}^{x}=\partial\omega_{\mathbf{q}}/\partial
q_{x}=gq_{x}/(m_{p}\omega_{\mathbf{q}})$ the group velocity in the tranport
$(x)$ direction. We may define
a Lorenz number%
\[
L_{d}\equiv\frac{\kappa}{\sigma T}=\frac{4m_{p}^{2}k_{B}^{4}T^{2}}{\hbar^{4}%
}\left[  \frac{\partial\ln\chi}{\partial P_{0}}\right]  ^{-2}\left\{
\begin{array}
[c]{c}%
\xi_{0}^{4},\\
\frac{64\pi^{3}}{45}\xi_{0},
\end{array}%
\begin{array}
[c]{c}%
\xi_{0}\gg 1\\
\xi_{0}\ll 1
\end{array}
\right.
\]
that is material specific and, assuming that the other parameters are
approximately constant, scales with $T^{-2}$ ($T$) at low (high) temperatures.

Next, we
consider a quasi-one dimensional ballistic ferroelectric wire that connects to
reservoirs. Within the linear response regime, the effective field ($\Delta
E$) and temperature ($\Delta T$) differences between the reservoirs generate
the polarization $(J_{p})$ and heat $(J_{q})$ currents as
\cite{tang2021ferroelectric}
\begin{equation}
\left(
\begin{matrix}
-J_{p}\\
J_{q}%
\end{matrix}
\right)  =G\left(
\begin{matrix}
1 & S_{b}\\
\Pi_{b} & K/G
\end{matrix}
\right)  \left(
\begin{matrix}
\Delta E\\
-\Delta T
\end{matrix}
\right)  ,
\end{equation}
noting that the currents driven by an effective field difference are
transient. The polarization ($G$) and thermal ($K$) conductances and the
ballistic Seebeck ($S_{b}$) and Peltier ($\Pi_{b}=TS_{b}$) coefficients follow
from the Landauer-B\"{u}ttiker formalism \cite{tang2021ferroelectric}:
\begin{align}
G &  =\frac{1}{\hbar}\int(\delta p_{k})^{2}\left(  -\frac{\partial f_{0}%
}{\partial\omega_{k}}\right)  \frac{d\omega_{k}}{2\pi}\nonumber\\
&  =\frac{\hbar\chi}{8\pi m_{p}}\left[  \frac{\partial\ln\chi}{\partial P_{0}%
}\right]  ^{2}\left\{
\begin{array}
[c]{c}%
e^{-\xi_{0}},\\
\frac{1}{3}\xi_{0}^{-1},
\end{array}%
\begin{array}
[c]{c}%
\xi_{0}\gg1\\
\xi_{0}\ll1
\end{array}
\right. \\
S_{b} &  =\frac{1}{\hbar(GT)}\int(-\delta p_{k})\hbar\omega_{k}\left(
-\frac{\partial f_{0}}{\partial\omega_{k}}\right)  \frac{d\omega_{k}}{2\pi
}\nonumber\\
&  =\frac{\hbar}{4\pi m_{p}(GT)}\frac{\partial\ln\chi}{\partial P_{0}}%
f_{0}\left( \xi_{0}\right) \\
K &  =\frac{1}{\hbar T}\int(\hbar\omega_{k})^{2}\left(  -\frac{\partial f_{0}%
}{\partial\omega_{k}}\right)  \frac{d\omega_{k}}{2\pi}\nonumber\\
&  =K_{0}\left\{
\begin{array}
[c]{c}%
\frac{3}{\pi^{2}}\xi_{0}^{2}e^{-\xi_{0}},\\
1,
\end{array}%
\begin{array}
[c]{c}%
\xi_{0}\gg1\\
\xi_{0}\ll1
\end{array}
\right.
\end{align}
where $k$ is the wave vector of the ferrons propagating along the wire with
the dispersion relation $\omega_{k}$, $K_{0}=\pi k_{B}^{2}T/(6\hbar)$ the
single-mode quantum thermal conductance and the summation over transverse modes
was restricted to the lowest subband.\textbf{\ }The Lorenz number turns out to
be quite different
\begin{equation}
L_{b}\equiv\frac{K}{GT}=\frac{4}{(T\chi)^{2}}\left[  \frac{\partial\ln\chi
}{\partial P_{0}}\right]  ^{-2}\left\{
\begin{array}
[c]{c}%
1,\\
\pi^{2}\xi_{0}^{-1},
\end{array}%
\begin{array}
[c]{c}%
\xi_{0}\gg1\\
\xi_{0}\ll1
\end{array}
\right.  .
\end{equation}

\begin{table}[ptb]
\caption{The material parameters introduced in the text for selected
perovskite ferroelectrics at room temperature.}%
\label{tab:table1}%
\begin{ruledtabular}
\begin{tabular}{cccccccc}
&BaTiO$_3$ \cite{hlinka2006phenomenological}& PbTiO$_3$
\cite{haun1987thermodynamic}& LiNbO$_3$ \cite{scrymgeour2005phenomenological} & units\\
\hline
$\alpha$& $-5.544\times 10^{-2} $ & $-0.3416$ & $-2.012$ & $10^{9}$\thinspace Jm/C$^{2}$
\\
$\beta$ & $-2.590$ & $-0.29$ & $3.608$ & $10^{9}$\thinspace Jm$^{5}$/C$^{4}$\\
$\lambda$ &$4.802$& $0.1563$ & $0$ & $10^{10}$\thinspace Jm$^9$/C$^{6}$\\
$g$ & $5.1$  & $2$ \cite{behera2011structure} & $5.39$ \cite{richman2019inadequacy} &$10^{-10}$Jm$^3$/C$^{2}$ \\
$m_{p}$& $1.35$ & $1.59$ \cite{morozovska2016influence}& $1.81$ & $10^{-18}$Jms$^2$/C$^{2}$ \\
$\tau$ \cite{tau} & 0.21 \cite{fontana1994quasimodes}  & 0.15 \cite{sanjurjo1983high} & 0.54 \cite{ridah1997temperature} & ps
\\
$V_{0}$ & 66 & 63.18 & 317.73& $\mathrm{\AA}^{3}$
\\
\end{tabular}
\end{ruledtabular}
\end{table}

All the above transport coefficients depend on an applied uniform electric
field via the field-dependence of $P_{0}$ (see Eq.~(\ref{spon})). The
integrand of the diffuse thermal conductivity Eq.~(\ref{DTR}) depends on the
field only via the occupation numbers,
\begin{align}
\kappa^{\prime}\equiv &  \frac{\partial\kappa}{\partial E}=\frac{\tau}%
{T\hbar^{2}}\int(v_{\mathbf{q}}^{x})^{2}(\hbar\omega_{\mathbf{q}})^{2}\delta
p_{\mathbf{q}}\frac{\partial^{2}f_{0}}{\partial\omega_{\mathbf{q}}^{2}}%
\frac{d^{3}\mathbf{q}}{(2\pi)^{3}}\nonumber\\
= &  -\sigma S_{d}\left\{
\begin{array}
[c]{c}%
\xi_{0},\\
3,
\end{array}%
\begin{array}
[c]{c}%
\xi_{0}\gg1\\
\xi_{0}\ll1
\end{array}
\right. \label{fKd}%
\end{align}
where the thermal conductance drops with a positive electric field along
$\mathbf{P}_{0}$ by electric \textquotedblleft freeze out\textquotedblright%
\ of the thermally excited ferrons. We also find
\begin{equation}
K^{\prime}\equiv\frac{\partial K}{\partial E}=-\xi_{0}\left[  1+f_{0}\left(  \xi_{0}\right)  \right]
GS_{b}.\label{fKb}
\end{equation}
Thus the $\kappa^{\prime}$ ($K^{\prime}$) together with the $L_{d}$ ($L_{b}$)
allows one to access $\sigma$ ($G$) and $S_{d}$ ($S_{b}$).

Table~\ref{tab:table2} summarizes the numerical calculations of the integral
expressions of transport coefficients derived above with the parameters given
in Table~\ref{tab:table1}, in which the integrals are cut-off by the Debye
wave vector $q_{D}=(6\pi^{2}/V_{0})^{1/3}$. We observe that the experimental
thermal conductivities are much larger than the computed ones because they are
dominated by the acoustic phonons and that the $\kappa^{\prime}$ and
$K^{\prime}$ agree well with the relations $\kappa^{\prime}\approx-3\sigma
S_{d}$ and Eq.~(\ref{fKb}), respectively.

\begin{table}[ptb]
\caption{The ferron gap ($\omega_{0}$) and dipole ($\delta p_{0}$) at the
$\Gamma$-point and transport coefficients for the ferroelectrics in Table
\ref{tab:table1} at room temperature, in which the field is at zero. The experimental total thermal
conductivities $(\kappa_{\mathrm{tot}}^{\text{\textrm{exp}}})$ are given for
comparison.}%
\label{tab:table2}%
\begin{ruledtabular}
\begin{tabular}{cccccccc}
&BaTiO$_3$ & PbTiO$_3$ & LiNbO$_3$ &  units\\
\hline
$\omega_{0}$& $20$ & $32$ & $47$ &  THz\\
$\delta p_{0}$& $-2.75$ & $-0.45$ & $-0.15$ &  e\AA
\\
$\sigma$ & $1.0$ & $3.4\times 10^{-2}$ & $7.8\times 10^{-3}$ & $10^{-15}$\thinspace m/$\Omega$\\
$G$ & $1.72$ & $2.7\times 10^{-2}$ & $1.9\times 10^{-3}$ & $10^{-24}$\thinspace m$^2$/$\Omega$ \\
$S_{d}$ &$0.16 $ & $0.72$ & $1.94 $ & $10^{7}$\thinspace V/(Km)\\
$S_{b}$ &$0.04 $ & $0.32$ & $1.18 $ & $10^{7}$\thinspace V/(Km)\\
$\kappa$ & $2.03$  & $0.74$ &$1.02$ & W/(Km) \\
$K$ & $0.75$  & $0.47$ &$0.34$ & $K_0$\\
$\kappa^{\prime}$& $-4.99$ & $-0.67$ & $-0.42$& 10$^{-9}$\thinspace W/(KV) \\
$K^{\prime}/K_{0}$& $-3.11$ & $-0.42$ & $-0.12$& $10^{-9}$\thinspace m/V \\
$\kappa_{\text{tot}}^{\text{exp}}$ & 6.5 \cite{suemune1965thermal} &  3.9 \cite{langenberg2019ferroelectric} & 8.5 \cite{burkhart1977determination} & W/(Km) \\
\end{tabular}
\end{ruledtabular}
\end{table}

The ferron dipole in BaTiO$_{3}$ is about 6 times (one order of magnitude)
larger than in PbTiO$_{3}$ (LiNbO$_{3}$) because of a larger anharmonicity
($\beta$ and $\lambda$) relative to the quadratic coefficient ($\alpha$) in
Eq.~(\ref{emp}). Hence, the polarization transport coefficients and the field
derivative of the thermal conductivity (conductance) $\kappa^{\prime}$
($K^{\prime}$) are largest in BaTiO$_{3}$. A \emph{negative} $\kappa^{\prime}$
can provide evidence for ferronic transport \cite{heremans}. However, when
comparing with experiments several competing mechanisms should be considered.
While to leading order acoustic phonons do not carry an electric dipole, the
electric field also modulates the elastic parameters including the sound
velocities by electrostriction and thereby heat transport, which could be
separated in prinicple by clamping the sample. A second order effect of the
electrostriction is a dynamical coupling of the acoustic phonons with the
ferrons that preserves $\kappa^{\prime}<0$ at low temperatures \cite{PT}.
Finally, electric fields suppress domain walls, which leads to an
\emph{increasing} thermal conductivity via a field-dependent relaxation time
\cite{mante1971phonon,northrop1982phonon,weilert1993mode,
langenberg2019ferroelectric}.

\begin{figure}[ptb]
\centering
\par
\includegraphics[width=5.2 cm]{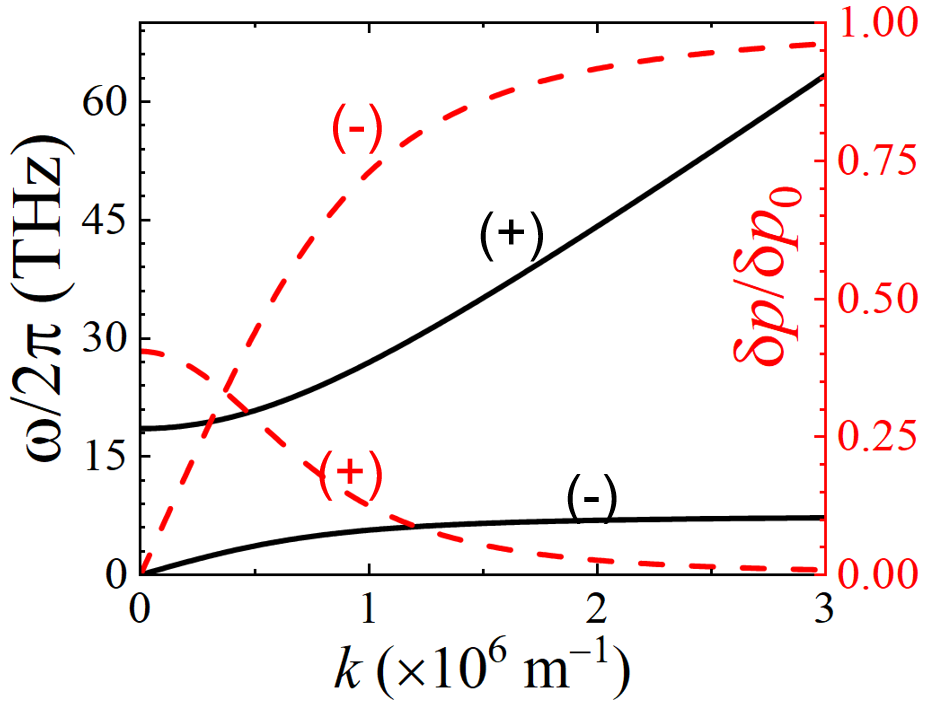}\newline\caption{The dispersion
relations (solid black curves) and the corresponding electric dipoles (dashed
red curves) of two $(\pm)$ of ferron polaritons branches in the absence of damping. The electric dipoles
are normalized by the value at the origin $\delta p_{0}=-0.15$\thinspace
e\AA . The parameters are for LiNbO$_{3}$ with $\varepsilon(\infty)=5.5$. }%
\label{Fig-2}%
\end{figure}

\emph{Electric dipole of ferron polaritons.} Photons can hybridize with
optical phonons to form phonon polaritons
\ \cite{born1955dynamical,fano1956atomic,henry1965raman,
bakker1992observation,kojima2002terahertz,kojima2003far,ikegaya2015real}, that
can show anharmonicities in ferroelectrics \cite{bakker1994investigation,
bakker1998coherent}. We may therefore consider \textquotedblleft ferron
polaritons\textquotedblright\ with a dispersion relation governed by
\cite{born1955dynamical}
\begin{equation}
\frac{c^{2}k^{2}}{\omega^{2}}=\varepsilon(\omega)\label{polariton}%
\end{equation}
where $c$, $k$ and $\varepsilon(\omega)$ are the light velocity, wave vector
and the dynamic (relative) permittivity in the long-wavelength limit,
respectively. According to Eq.~(\ref{hfl})
\begin{equation}
\varepsilon(\omega)-\varepsilon(\infty)\equiv\frac{\delta P_{\mathrm{h}}%
}{\varepsilon_{0}E_{\mathrm{th}}}=\frac{1}{m_{p}\varepsilon_{0}(\omega_{0}%
^{2}-\omega^{2}-i\omega\Tilde{\gamma})}%
\end{equation}
where $\tilde{\gamma}=\gamma/m_{p}$, while $\varepsilon(\infty)$ is the
high-frequency permittivity. While this dispersion is identical to that of the
phonon polaritons in normal ionic crystals
\cite{born1955dynamical,fano1956atomic}, the ferron polaritons may transport
electric dipoles below $T_{c}$. By Eq.~(\ref{emp2}), the electric dipole of
ferron polaritons reads
\begin{equation}
\delta p_{\pm}(k)=-\left.  \frac{\partial\hbar\omega_{\pm}(k)}{\partial
E}\right\vert _{E\rightarrow0}=\frac{\partial\omega_{\pm}(k)}{\partial
\omega_{0}}\delta p_{0}%
\end{equation}
where $+(-)$ indicates two (optical and ferronic) branches and $\delta
p_{0}=-\partial\hbar\omega_{0}/\partial E$. Figure~\ref{Fig-2} gives the
dispersion relations and the electric dipoles carried by the two branches for
LiNbO$_{3}$, in which the level repulsion renders the dipole of the ferronic
branch smaller than $\delta p_{0}$ even at $k=0$. Focused optical excitations
at the optical phonon frequency of ferroelectrics can therefore be a source of
coherent polarization currents and give rise to unique electrooptic properties
such as electric field-controlled light propagation. Electric-dipolar
interaction importantly affects the surface ferron-polariton dispersion
relations \cite{Rezende}.

\emph{Conclusions:} We identify the quasi-particle excitations of displacive
ferroelectrics that carry heat and electric dipole currents and predict the
associated low-temperature pyroelectric or electrocaloric coefficients, the
(field-dependent) thermal conductivity, Peltier and Seebeck coefficients, and
ferron polariton polarization. Thermally driven and electrically tunable
ferronic transport in a broad class of ferroelectric materials may provide
unique functionalities to thermal management and information technologies.

\emph{Acknowledgements:} We are grateful for enlightening discussions with
Beatriz Noheda, Bart J. van Wees, Joseph P. Heremans, and Sergio Rezende. JSPS
KAKENHI Grant No. 19H00645 supported P.T. and G.B. R.I. and K.U. acknowledge
support by JSPS KAKENHI Grant No. 20H02609, JST CREST \textquotedblleft
Creation of Innovative Core Technologies for Nano-enabled Thermal
Management\textquotedblright Grant No. JPMJCR17I1, and the Canon Foundation.

\bibliography{Refer}

\begin{thebibliography}{70}
\bibitem [{tau()}]{tau}%
  \BibitemOpen
  \href@noop {} {\bibinfo  {journal} {$\tau\sim 1/\Delta\omega$ was estimated
  from the broadening ($\Delta\omega$) of the lowest A$_1$(TO) line in
  Raman-scattering spectra}\ }\BibitemShut {NoStop}%
\end{thebibliography}
\end{document}